\documentclass[preprint,showpacs,showkeys,preprintnumbers,amsmath,amssymb]{revtex4}

\usepackage{natbib}
\usepackage{graphicx}
\usepackage{dcolumn}
\usepackage{bm}
\usepackage{longtable}

\begin{document}


\title{Global minima of Al$_{N}$, Au$_{N}$ and Pt$_{N}$, $N \le 80$, clusters described
by Voter-Chen version of embedded-atom potentials}

\author{Ali Sebetci}
 \email{asebetci@cankaya.edu.tr}
\affiliation{Department of Computer Engineering, \c{C}ankaya
University, 06530 Balgat Ankara, Turkey}

\author{Ziya B. G\"{u}ven\c{c}}
\email{guvenc@cankaya.edu.tr} \affiliation{Department of
Electronic and Communication Engineering, \c{C}ankaya University,
06530 Balgat Ankara, Turkey}

\date{\today}

\begin{abstract}
Using the basin-hopping Monte Carlo minimization approach we
report the global minima for aluminium, gold and platinum metal
clusters modelled by the Voter-Chen version of the embedded-atom
model potential containing up to 80 atoms. The virtue of the
Voter-Chen potentials is that they are derived by fitting to
experimental data of both diatomic molecules and bulk metals
simultaneously. Therefore, it may be more appropriate for a wide
range of the size of the clusters. This is important since almost
all properties of the small clusters are size dependent. The
results show that the global minima of the Al, Au and Pt clusters
have structures based on either octahedral, decahedral,
icosahedral or a mixture of decahedral and icosahedral packing.
The 54-atom icosahedron without a central atom is found to be more
stable than the 55-atom complete icosahedron for all of the
elements considered in this work. The most of the Al global minima
are identified as some fcc structures and many of the Au global
minima are found to be some low symmetric structures, which are
both in agreement with the previous experimental studies.
\end{abstract}

\maketitle

\section{\label{sec:level1}INTRODUCTION}

Since Richard Feynman's famous challenging talk \textit{There's
Plenty of Room at the Bottom} in 1959~\cite{Feynman}, many
scientists all over the world are still studying on the
investigation and fabrication of nanometer scale ($10^{-9}$ m)
structures and devices. In his talk, he challenged scientists to
develop a new field of study where devices and machines could be
constructed from components consisting of a small number (tens or
hundreds) of atoms. The use of metal and semiconductor clusters as
components of nanodevices is one of the most important reasons
which explains why there are considerable theoretical and
experimental interest in the study of gas phase and supported
metal clusters in the last few
decades~\cite{Haberland,Schmid,Martin,Jellinek,Johnston}. Due to
their finite size, these small particles may have totally
different structures and material properties than their bulk
crystalline forms. Furthermore, these properties may sometimes
change drastically whenever a single atom is added to or removed
from the cluster~\cite{Eberhardt}. A systematic study of evolution
of these properties with size allows elucidation of the transition
from the molecular structure to condensed matter phase. Clusters,
in particular metal clusters, play an important role in many
chemical reactions as catalysts, as well. The structure of small
metal clusters in a reaction can have a major effect on the rate
of formation of products~\cite{Jel-Guv}.

In this study, using the basin-hopping~\cite{Wales} Monte Carlo
minimization approach we report the global minima for aluminium,
gold and platinum metal clusters modelled by the
Voter-Chen~\cite{Voter} version of the embedded-atom model
(EAM)~\cite{Baskes} potential containing up to 80 atoms. The
virtue of the Voter-Chen potentials is that they are derived by
fitting to experimental data of both diatomic molecules and bulk
metals simultaneously. Therefore, it may be more appropriate for a
wide range of the size of the clusters. This is important since
almost all properties of the small clusters are size dependent.

This paper is organized as follows: The interaction potential and
the computational procedure will be discussed in Section II.
Results and discussions are presented in Section III, and
conclusions are given in Section IV.

\section{\label{sec:level1}COMPUTATIONAL METHODS}

\subsection{The Voter-Chen Potential}

In any $N$-scaling energy expression, the total energy, $E_{tot}$
of a system of $N$ atoms can be written as a sum
\begin{equation}
E_{tot}=\sum_{i}^{N}E_{i}.
\end{equation}
In the EAM, the configuration energy $E_{i}$ of each atom $i$ is
represented as
\begin{equation}
E_{i}=\frac{1}{2}\sum_{j \ne
i}\phi_{ij}(r_{ij})+F_{i}(\bar{\rho_{i}}),
\end{equation}
where $F_{i}$ is the embedding term, $\phi_{ij}$ is the
pairwise-addition part of the interaction between atoms $i$ and
$j$, $r_{ij}$ is the distance between atoms $i$ and $j$, and
$\bar{\rho_{i}}$ is the total "host" electron density at the
position of atom $i$:
\begin{equation}
\bar{\rho_{i}}=\sum_{j \ne i}\rho_{j}(r_{ij}).
\end{equation}
The sums over neighboring atoms $j$ are limited by the range of
the cutoff for $\phi$ and $\rho$, which is approximately 5 $\AA$
for the metals considered in this work. Key to the EAM is the
nonlinearity of the function $F(\bar{\rho})$ which provides a
many-body contribution to the energy. If $F$ were purely linear,
the two terms in Eq.2 could be collapsed to give a simple pair
potential. Thus, a nonlinear $F(\bar{\rho})$ provides a many-body
contribution to the energy. Because $\bar{\rho_{i}}$ depends only
on scalar distances to neighboring atoms, the many-body term has
no angular dependence. Nonetheless, this spherically symmetric,
many-body interaction is quite important.

All the parameters in the Voter and Chen model were determined by
minimizing the root-mean-square deviation ($\chi_{rms}$) between
the calculated and experimental values of three elastic constants
($C_{11}$, $C_{12}$, and $C_{44}$), the unrelaxed vacancy
formation energy ($E^{f}_{vac}$) of the bulk metals (Al, Au and
Pt), and of the bond length ($R_{e}$) and bond strength ($D_{e}$)
of their diatomic molecules.

\subsection{The Basin-Hopping Algorithm}

Two new and more successful algorithms have been developed within
the last two decades to search the global minimum of an energy
landscape, which are different than the traditional random search
and simulated annealing techniques: basin-hopping and genetic
algorithms. The genetic algorithm is a search based on the
principles of natural evolution~\cite{Goldberg}, while the
basin-hopping approach belongs to the family of hypersurface
deformation methods~\cite{Stillinger} where the energy is
transformed to a smoother surface. The basin-hopping algorithm
which we have used in the present work is based upon Li and
Scheraga's~\cite{Li} Monte Carlo (MC) minimization, and it has
been developed and employed for several systems by Doye and
Wales~\cite{Wales,Doye1,Doye2}. In the basin-hopping algorithm,
the transformed potential energy surface (PES),
$\tilde{E}(\bold{X})$, is defined by $\tilde{E}(\bold{X})=
min\{E(\bold{X})\}$, where $\bold{X}$ represents the vector of
atomic coordinates and $min$ signifies that an energy minimization
is performed starting from $\bold{X}$. Unlike many PES
transformations, this basin-hopping transformation guarantees to
preserve the identity of the global minimum. The topography of the
transformed surface is that of a multi-dimensional staircase ( a
set of interpenetrating staircases with plateaus corresponding to
the basins of attraction of each minimum). Since the barriers
between the local minima are removed in the transformed PES,
vibrational motions within the well surrounding a minimum are
removed. In addition, transitions from one local minimum to
another in the transformed PES can occur at any point along the
boundary between these local minima, whereas on the untransformed
surface transitions can occur only when the system passes through
the transition state. Consequently, on $\tilde{E}(\bold{X})$, the
system can hop directly between the basins; hence it is the name
of this transformation.

We have used the GMIN~\cite{opt} program in our simulations to
locate the lowest energy structures of the Voter-Chen Al, Au and
Pt clusters. The MC runs have been started with the configurations
which are the global minima of the Morse clusters. For a given
size, as the interaction range of the Morse potential changes, the
global minimum varies. Different global minima for different
interaction ranges of the Morse potential were reported up to
80-atom clusters before~\cite{Doye3,Doye4}. We have reoptimized
all these Morse global minima by performing several MC runs of
100,000 steps of each.

\section{\label{sec:level1}RESULTS AND DISCUSSION}

\subsection{Aluminium Clusters}

It goes back to the middle of the 1980s that a number of
theoretical studies of Al clusters have been carried out by
different
groups~\cite{Chou,Jug,Pacchioni,Upton,Pettersson,Cheng,Yi,Jones,
Akola,Ahlriches,Khanna,Gong,Jellinek2,Rao,Erkoc,Johnston2,Lloyd1,Lloyd2,Turner,Joswig}.
These studies range from the simple jellium model~\cite{Chou}
where the cluster geometry is ignored, to a number of models where
the geometry explicitly enters into the picture including
semiempirical molecular orbital calculations~\cite{Jug}, quantum
molecular dynamics~\cite{Yi,Jones,Akola,Ahlriches,Gong},
quantum-mechanical calculations based on
quantum-chemical~\cite{Pacchioni,Upton,Pettersson} and
density-functional~\cite{Cheng,Yi,Jones,
Akola,Ahlriches,Khanna,Gong,Jellinek2,Rao} theories (DFT) within
local density or local spin-density approximations, molecular
dynamics and Monte Carlo simulations based on empirical model
potentials~\cite{Erkoc,Johnston2,Lloyd1,Lloyd2,Turner,Joswig}.
Especially the icosahedral Al$_{13}$ has been studied
intensively~\cite{Pettersson,Jellinek2}. The most recent and more
extensive density-functional calculations have been presented by
Ahlrichs and Elliott~\cite{Ahlriches} and by Rao and
Jena~\cite{Rao} in 1999. These studies focused both on electronic
and structural properties of neutral and ionized Al clusters up to
15 atoms, respectively. On the other hand, while the empirical
model potential
studies~\cite{Erkoc,Johnston2,Lloyd1,Lloyd2,Turner,Joswig} cannot
calculate the electronic properties of the clusters, it is
possible to search PES of higher sized clusters with them since
they are computationally much less demanding than {\it ab initio}
calculations. In these model potential studies carried out by
random search, simulated annealing or genetic algorithms, Al
clusters are described by an empirical many-body
potential~\cite{Erkoc}, two-plus-three body Murrell-Mottram
potential~\cite{Johnston2,Lloyd1,Lloyd2}, Gupta~\cite{Turner} or
Sutton-Chen~\cite{Joswig} potentials. Similarly, the experimental
studies on Al
clusters~\cite{Cox,Jarrold,Hanley,Saunders,Leuchtner,
Schriver,Heer,Gantefor,Taylor,Nakajima,Wu,Martin2} go back to the
middle of the 1980s. It is known that while the electronic factors
determine cluster stability for alkali metal
clusters~\cite{Knight}, packing and surface energy effects
dominate on the structure of alkaline earth elements, such as
calcium and strontium~\cite{Martin2}. Aluminium places at a
central position between the regimes of electronic and geometric
shells~\cite{Schriver}. Martin's mass spectroscopic
studies~\cite{Martin2} have shown that Al clusters with up to a
few hundred atoms have face-centred cubic (fcc) packing
structures. These experimental interpretations have been confirmed
by theoretical calculations using empirical
potentials~\cite{Turner} and DFT~\cite{Ahlriches}. Jarrold and
Bower have performed experiments on smaller Al clusters which
enabled them to determine the topologies of clusters with tens of
atoms~\cite{Jarrold}.

\begin{figure}
\includegraphics[bb= 0 0 24.55cm 35.88cm, scale=0.55]{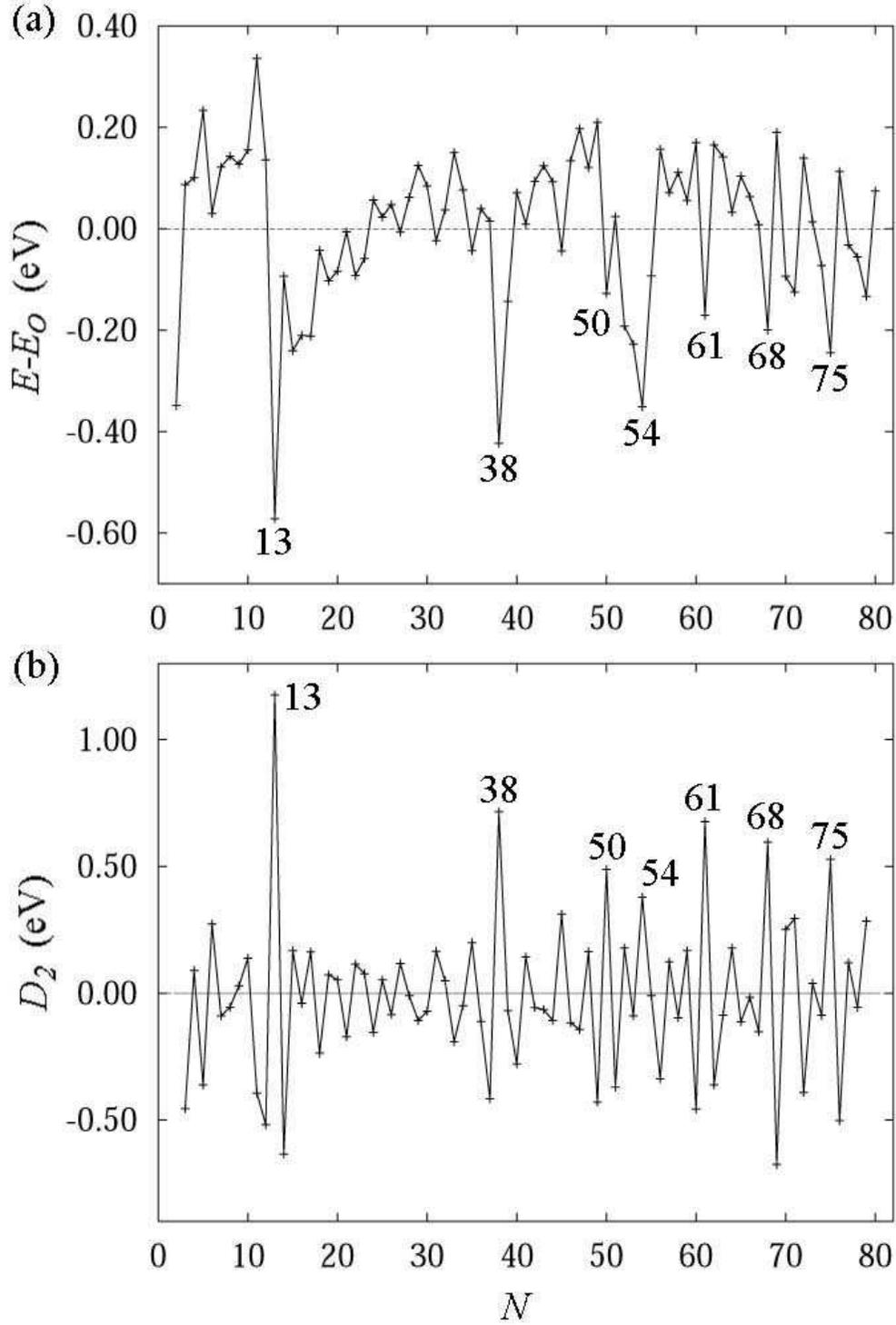}
\caption{\label{energy-Al} (a) $E-E_{0}$ is the relative energies
of quenched Al clusters where
$E_{0}=5.09182-2.96861N^{1/3}+2.7261N^{2/3}-3.43728N$; (b) The
second finite difference in binding energy v.s. size $N$. }
\end{figure}

We have reported the total energies ($E$), the point groups
($PG$), and the structural assignments ($SA$) (whenever possible)
of the global minima for the Al clusters up to 80 atoms described
by the Voter-Chen potential in Table~\ref{table-Al}. The point
groups of the structures are determined with OPTIM
program~\cite{opt}. Symmetry elements are diagnosed when rotation
and reflection operators produce the same geometry (correct to
0.001) in each Cartesian coordinates. The energies and the second
finite differences in energies
\begin{equation}
D_{2}E(N)=E_{l}(N+1)+E_{l}(N-1)-2E_{l}(N)
\end{equation}
are plotted in Figs.~\ref{energy-Al}(a) and (b), respectively.
Following Northby \textit{et al.}~\cite{Northby} and Lee and
Stein~\cite{Stein}, the function,
\begin{equation}
E_{0}=aN + bN^{2/3} + cN^{1/3} + d ,
\end{equation}
is fitted to the energies given in Table~\ref{table-Al}, and it is
subtracted from the energies of the clusters in order to emphasize
the size dependence. In this polynomial function, $a$ describes
the volume, $b$ surface, $c$ edge, and $d$ the vertex
contributions to the energy. $D_{2}E$ is generally correlated with
the magic numbers observed in mass spectra. Clusters are
particularly abundant at magic number sizes in mass spectra since
they are the most stable ones~\cite{Clemenger}.

\begin{figure}
\includegraphics[bb= 0 0 14.36cm 22.63cm, scale=1.0]{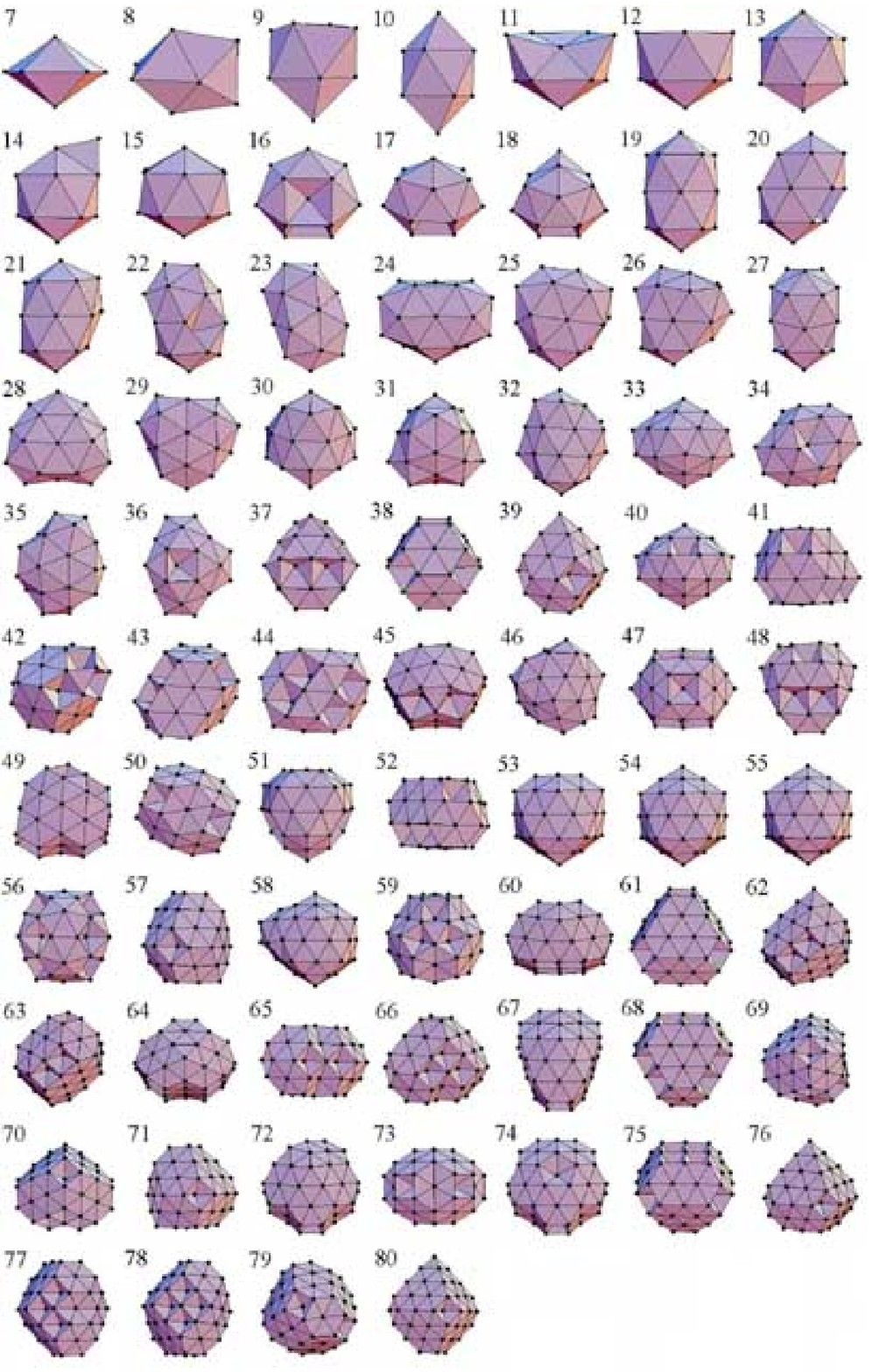}
\caption{\label{global-Al} Structures of the global minima for
$Al_{7}-Al_{80}$ clusters }
\end{figure}

The triangulated polyhedral structures of the Al$_{7}$-Al$_{80}$
global minima are illustrated in Fig.~\ref{global-Al}. The
structures for the first seven Al$_{N}$ clusters ($N=2-8$) are
similar to those obtained by other empirical potentials for
aluminum~\cite{Lloyd1,Joswig} and other metals~\cite{Doye1,Ali1}.
Al$_{3}$ forms an equilateral triangle, Al$_{4}$ a tetrahedron,
Al$_{5}$ a trigonal bipyramid, Al$_{6}$ an octahedron, Al$_{7}$ a
pentagonal bipyramid, and Al$_{8}$ is a bicapped octahedron. All
of these structures are located as the global minima of Au and Pt
clusters in the present work, too. Al$_{9}$ can be described as a
three capped trigonal prisms and Al$_{10}$ is a hexadecahedron,
which are the same with Joswig and Springborg's calculations of Al
clusters employed by Sutton-Chen potential~\cite{Joswig}.
Structures of the Al clusters with $N=11-14$ atoms are
icosahedral. The Al$_{15}$ is the sixfold icositetrahedron. The
16- and 17-atom Al clusters involve a mixture of decahedral and
icosahedral staking sequences. The Al$_{19}$ is a double
icosahedron. In the size range of $N=20-36$, all clusters have
face-sharing icosahedral (fsI) structures possessing generally low
symmetries. Above the size of 36, the most of the Al clusters are
fcc packed. This is consistent with Martin's experimental
study~\cite{Martin2} with the exceptions of the 40-, 51-, 53- and
54-atom uncentred icosahedral (ucI) structures, the 55-atom
centred icosahedron, and the 60-, 64-, 67-, 72-, 73- and 74-atom
decahedral (dec) structures. As a result the total number of fcc
Al clusters having more than 36 atoms is 26.

It can be seen from both of the Figs.~\ref{energy-Al}(a) and (b)
that the most stable structure occurs at size 13 which corresponds
to complete Mackay icosahedra~\cite{Mackay}. The other relatively
more stable structures with respect to their neighboring sizes are
$N$=38, 50, 54, 61, 68 and 75 corresponding to truncated
octahedron, twinned truncated octahedron, uncentred
icosahedra~\cite{Ali2}, and some other three fcc structures,
respectively.

\subsection{Gold Clusters}

Gold nanoparticles are a fundamental part of recently synthesized
novel nanostructured materials and
devices~\cite{Whetten,Mirkin,Andres}. Structural characterization
using a variety of experimental techniques can be performed on Au
clusters~\cite{Cleveland,Schaaff,Koga,Palpant,Spasov}. Experiments
suggest that gold nanoclusters with diameters of 1-2 nm,
corresponding to aggregates with $N$=20-200 atoms, are
amorphous~\cite{Cleveland,Schaaff}. The theoretical studies on
gold nanoclusters change from empirical MD or MC simulations using
EAM~\cite{Luedtke}, Gupta~\cite{Garzon}, Sutton-Chen~\cite{Doye1}
and Murrell-Mottram~\cite{Wilson} potentials to some
first-principle calculations using DFT~\cite{Haberlen,Wang},
generalized gradient approximation~\cite{Hakkinen}, spin-polarized
Becke-Lee-Yang-Parr functional~\cite{Gronbech}, and Hartree-Fock
and post Hartree-Fock levels~\cite{Bravo}.

\begin{figure}
\includegraphics[bb= 0 0 24.55cm 35.88cm, scale=0.55]{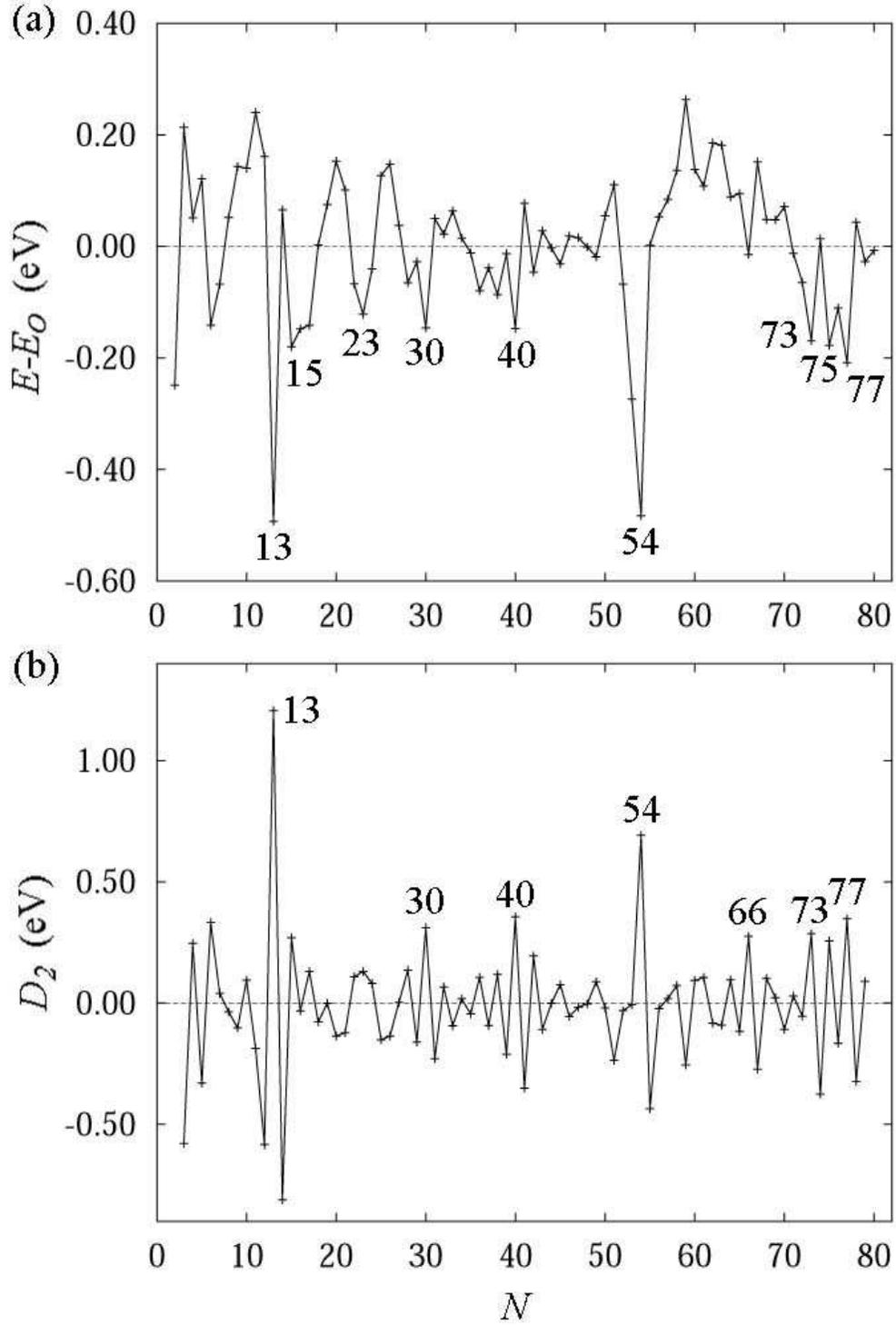}
\caption{\label{energy-Au} (a) $E-E_{0}$ is the relative energies
of quenched Au clusters where
$E_{0}=8.63706-6.88748N^{1/3}+3.97967N^{2/3}-4.15816N$; (b) The
second finite difference in binding energy v.s. size $N$. }
\end{figure}

\begin{figure}
\includegraphics[bb= 0 0 14.36cm 22.63cm, scale=1.0]{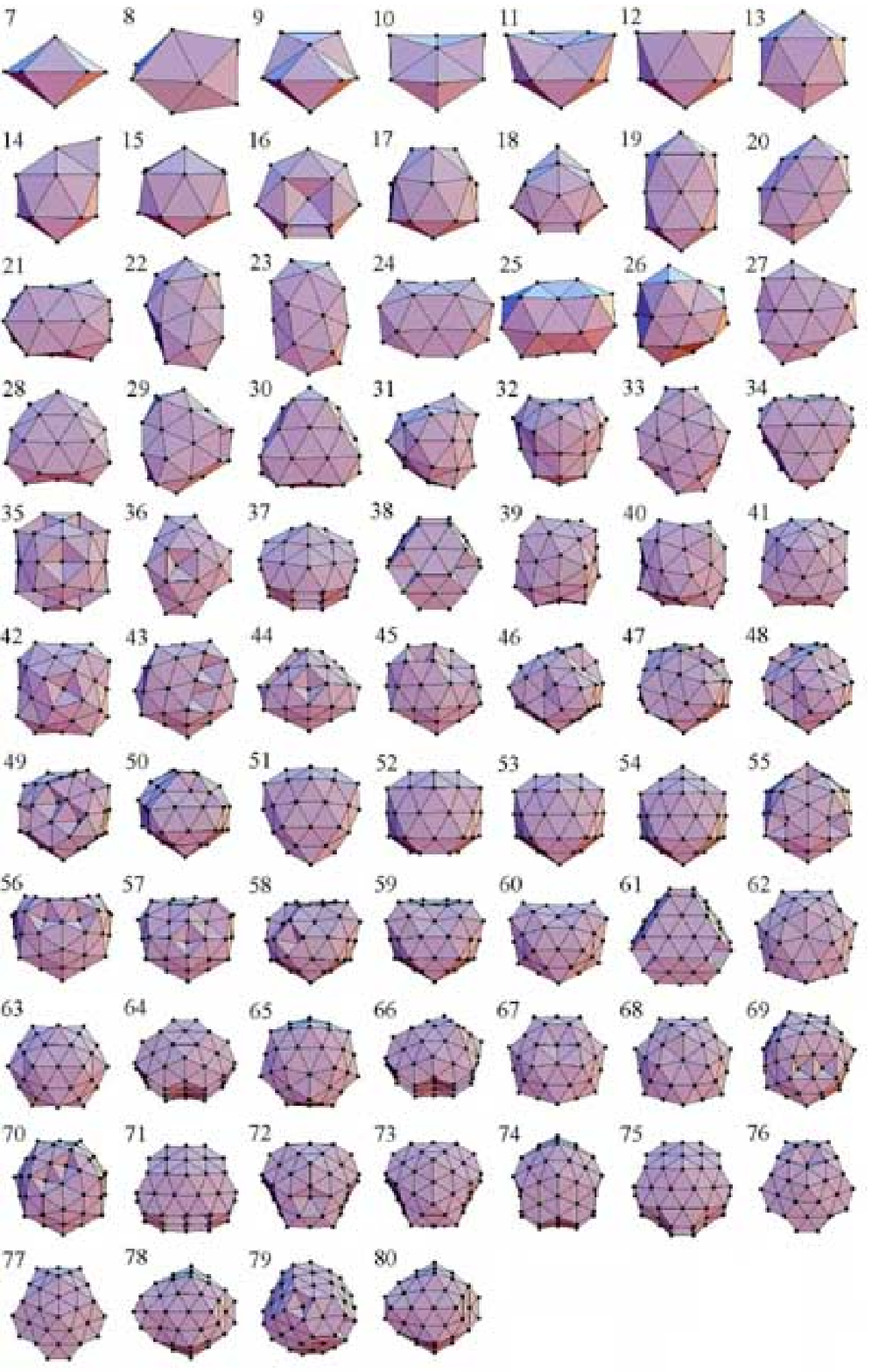}
\caption{\label{global-Au} Structures of the global minima for
$Au_{7}-Au_{80}$ clusters }
\end{figure}

We have reported the total energies ($E$), the point groups
($PG$), and the structural assignments ($SA$) (whenever possible)
of the global minima for the gold clusters of $N$=2-80 atoms
described by the Voter-Chen potential in Table~\ref{table-Au}. The
energies and the second finite differences in energies are plotted
in Figs.~\ref{energy-Au}(a) and (b), respectively. The
triangulated polyhedral structures of the Au$_{7}$-Au$_{80}$
global minima are illustrated in Fig.~\ref{global-Au}. In our
calculations we have found that Au$_{9}$-Au$_{14}$ clusters are
icosahedral. The 13-atom icosahedron has been reported as the
lowest energy structure of a Au$_{13}$ cluster by some of the
previous empirical studies~\cite{Doye1,Wilson} as well, although
they have presented some other structures for some of the gold
clusters in this size range. However, the icosahedron is not the
global minimum in the first principle calculations of Wang et
al.~\cite{Wang}. In addition, in many of the {\it ab initio}
studies the lowest energy structures of the small clusters are
found to be some planar forms~\cite{Haberlen,Wang,Hakkinen}. This
is because of the fact that since the empirical many body methods
are lack of directionality, these potentials favor more compact,
spherically symmetric structures. However, this discrepancy
between the first principle and empirical methods vanishes when
the cluster size increases. In our results the global minima of
Au$_{15}$, Au$_{16}$, Au$_{18}$, and Au$_{19}$ are the same as
those of the corresponding Al clusters. Similar to the Al
clusters, in the size range of $N=20-36$, all gold clusters have
fsI structures. The 37-atom cluster has a mixture of decahedral
and icosahedral morphologies. The 38-atom cluster is a truncated
octahedron. We have found only two more fcc structures (at $N=61$
and $N=79$) in the global minima of Au clusters above this size.
In agreement with many of the previous theoretical calculations,
the Au$_{55}$ is not a icosahedron in our calculations too,
although 52-, 53-, and 54-atom Au clusters are ucI. For the size
range of $N=64-79$, the dominant structural motif is the
decahedral morphology. While the 64-, 71-, and 75-atom clusters
have perfect decahedral structures, the 66-, 72-, 73-, 74-, 76-,
and 77-atom clusters have some icosahedral deficiencies on their
decahedral backbones. Our results for the Au clusters are in
agreement with the experimental suggestion that gold nanoclusters
with $N$=20-200 atoms are amorphous~\cite{Cleveland,Schaaff} since
the most of the structures reported in the present work have low
symmetry (i.e., $C_{s}$). Fig.~\ref{energy-Au}(b) suggests that
the most stable structures occur at sizes of 13, 30, 40, 54, 66,
73, 75 and 77. The 38-atom truncated octahedron does not seem as a
magic number of the Au clusters, instead a 40-atom amorphous
structure is more stable. For the higher sizes, decahedral
structures and mixtures of decahedral and icosahedral staking
sequences become more stable than the others, except the 54-atom
uncentred icosahedron.

\subsection{Platinum Clusters}

\begin{figure}
\includegraphics[bb= 0 0 24.55cm 35.88cm, scale=0.55]{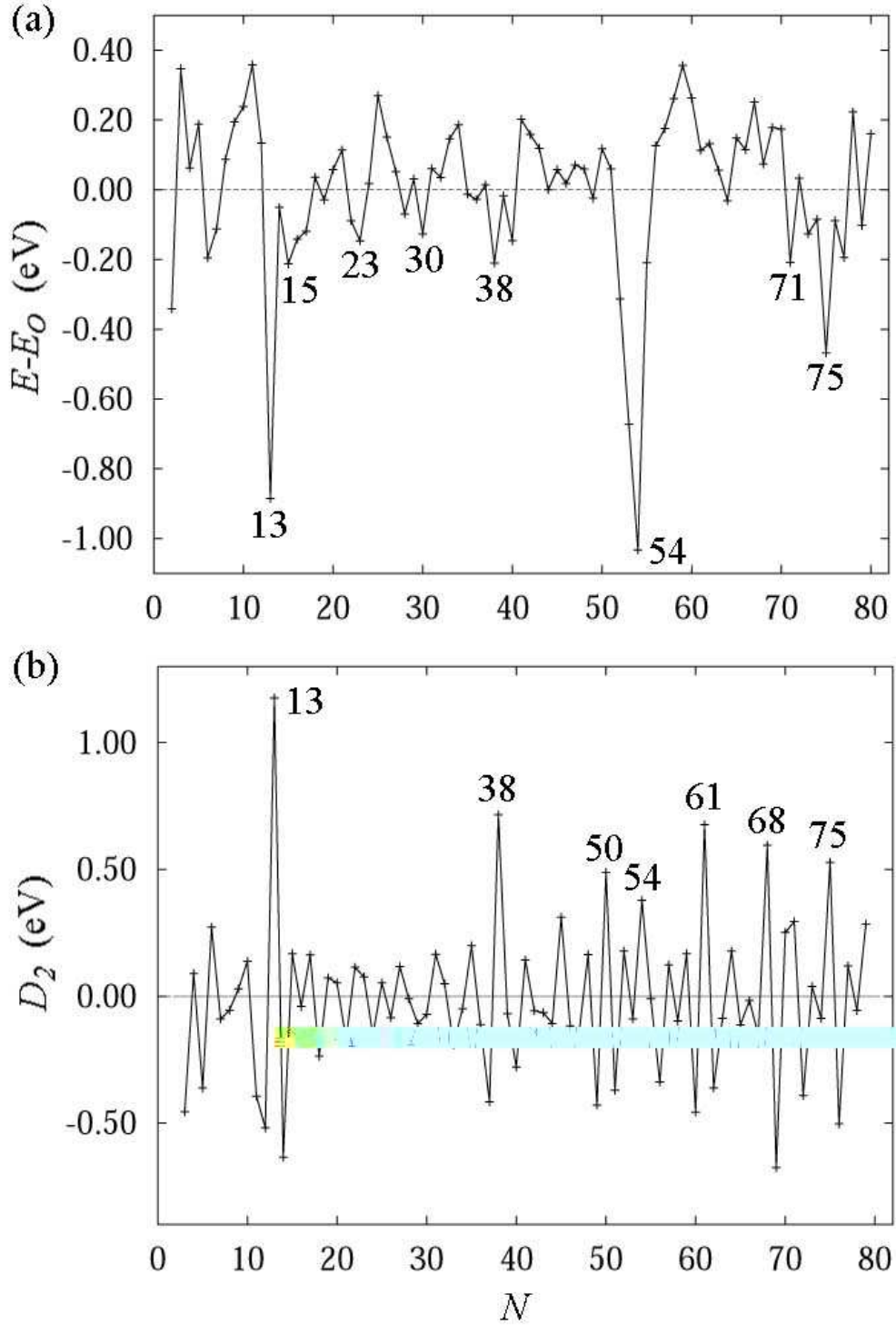}
\caption{\label{energy-Pt} (a) $E-E_{0}$ is the relative energies
of quenched Pt clusters where
$E_{0}=11.6998-9.27227N^{1/3}+5.88215N^{2/3}-6.08642N$; (b) The
second finite difference in binding energy v.s. size $N$. }
\end{figure}

\begin{figure}
\includegraphics[bb= 0 0 14.36cm 22.63cm, scale=1.0]{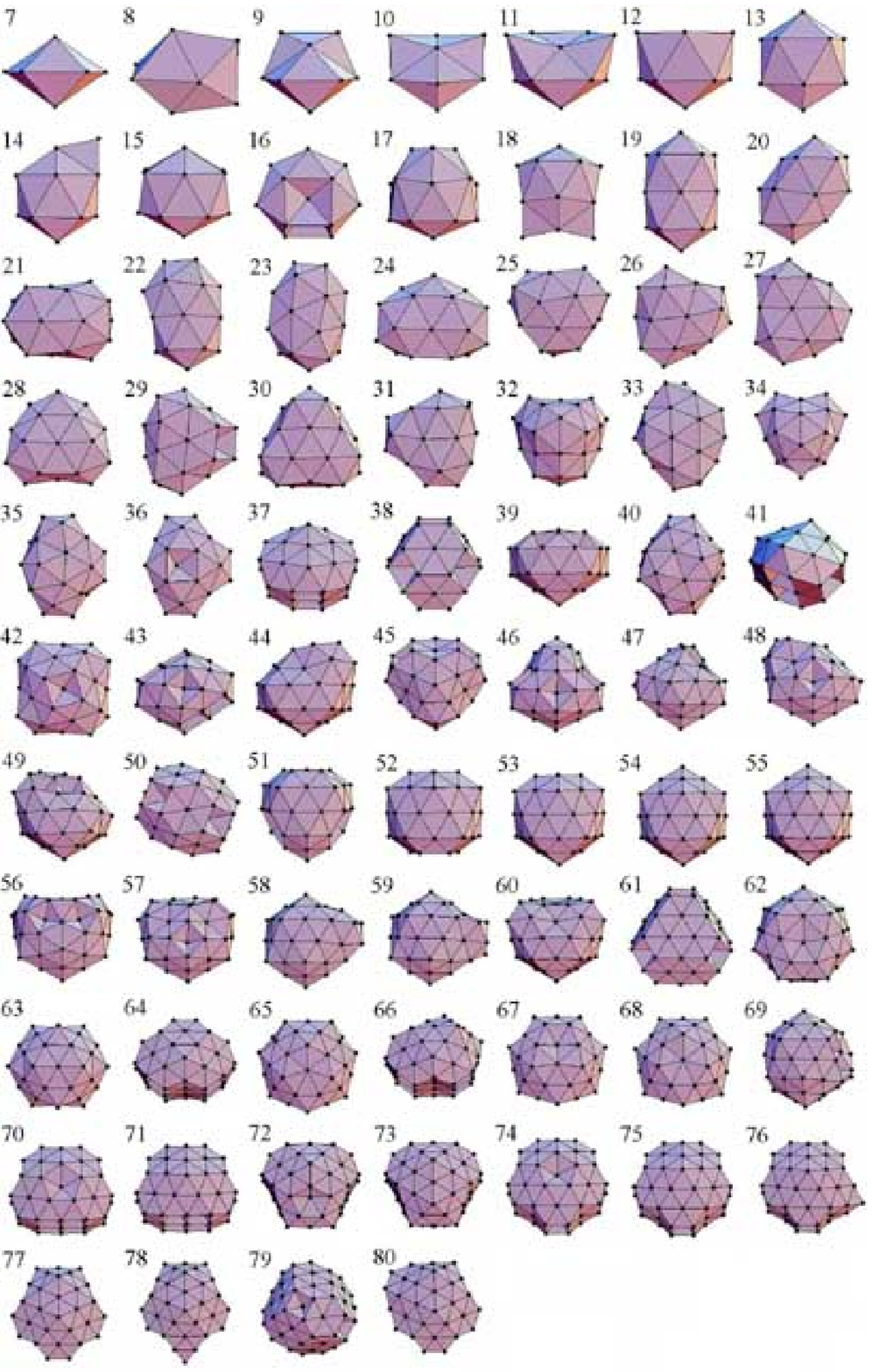}
\caption{\label{global-Pt} Structures of the global minima for
$Pt_{7}-Pt_{80}$ clusters }
\end{figure}

We have reported before the lowest energy structures, the numbers
of stable isomers, growth pathways, probabilities of sampling the
basins of attraction of the stable isomers, and the energy
spectrum-widths which are defined by the energy difference between
the most and the least stable isomers of Pt$_{2}$-Pt$_{21}$
clusters~\cite{Ali1} and the global minima of Pt$_{22}$-Pt$_{56}$
clusters~\cite{Ali2}. Since all relevant literature of platinum
clusters can be found in those studies, we do not repeat them here
once more. We have reported the total energies ($E$), the point
groups ($PG$), and the structural assignments ($SA$) of the global
minima of Pt clusters described by the Voter-Chen potential for $N
\le $ 80 atoms in Table~\ref{table-Pt}. The energies and the
second finite differences in energies are plotted in
Figs.~\ref{energy-Pt}(a) and (b), respectively. The triangulated
polyhedral structures of the Pt$_{7}$-Pt$_{80}$ global minima are
illustrated in Fig.~\ref{global-Pt}.

The lowest energy structures of the Pt clusters are more similar
to those of the Au clusters than those of the Al clusters. All the
global minima of Au and Pt clusters are identical for $N \le $ 17.
The 18-atom Pt cluster does not have the decahedral morphology of
the Au$_{18}$ cluster. In the size range of $N$=19-38, the most of
the Pt clusters have ucI structures which are similar to the cases
for both Al and Au clusters. In this size range, 12 Pt clusters
have identical structures with the corresponding Au clusters
(i.e., at the sizes of 19-21, 26, 28-30, 32, 33, 36-38). The main
differences between the Au and Pt clusters occur at the sizes of
41, 50, 51, 55, 70, 74, 76, 78, and 80: the 41-atom Pt cluster has
a mixture of decahedral and icosahedral morphologies, the 50-atom
Pt cluster is a twinned truncated octahedron, the 51-atom cluster
is an uncentred icosahedron missing three surface atoms, the
55-atom cluster is a complete Mackay icosahedron, the 70-, 74-,
and 76-atom clusters are some decahedrons and finally the 78- and
80-atom Pt clusters have a mixture of decahedral and icosahedral
staking sequences. For the higher sizes, while Pt clusters prefer
fully decahedral structures, the Au clusters favor structures
involving a mixture of decahedral and icosahedral staking
sequences (see the sizes of 70, 74, and 76). When the normalized
energy (Fig.~\ref{energy-Pt}(a)) and second finite difference in
energy plots (Fig.~\ref{energy-Pt}(b)) of the Pt clusters are
considered, it can be seen that the most stable sizes are 13, 38,
50, 54, 61, 68, and 75. Interestingly, these magic numbers are
more similar to those of the Al than those of the Au clusters.

\section{\label{sec:level1}CONCLUSIONS}

In the present study, we have reported the global minima of Al, Au
and Pt clusters up to 80 atoms described by the Voter-Chen version
of the EAM potential in a basin-hopping MC geometry minimization
technique. The results show that the global minima of the Al, Au
and Pt clusters have structures based on either fcc, decahedral,
icosahedral or a mixture of decahedral and icosahedral packing.
The 54-atom icosahedron without a central atom is found to be more
stable than the 55-atom complete icosahedron for all of the
elements considered in this work. The most of the Al global minima
are identified as some fcc structures as the previous experimental
studies suggest. Many of the Au global minima are found to be some
low symmetric structures, which is also in some agreement with the
experimental studies of the Au clusters. Although many of the Pt
global minima are identical with the global minima of the
corresponding Au clusters, the most stable sizes of the Pt
clusters occur at the same sizes of the Al clusters.


\begingroup
\squeezetable
\begin{table}
\caption{\label{table-Al}Global minima for Al clusters. For each
minimum energy ($E$), point group ($PG$) and structural assignment
($SA$) are given if possible. The structural categories are:
centred (cI), uncentred (ucI) and face-sharing icosahedral (fsI);
face centred cubic packed (fcc); decahedral with \textit{n} atoms
along the decahedral axis (dec(\textit{n})); involving a mixture
of staking sequences (mix).}
\begin{ruledtabular}
\begin{tabular}{l l l l l l l l}
$N$& $E$ (eV) & $PG$ & $SA$ & $N$ & $E$ (eV) & $PG$ & $SA$ \\
\hline
   &            &               &     & 41 &  -113.6500 & $C_{3v}$& fcc \\
2  &  -1.5443   & $D_{\infty h}$&     & 42 &  -116.5605 & $C_{s}$ & fcc \\
3  &  -3.7442   & $D_{3h}$      &     & 43 &  -119.5276 & $C_{s}$ & fcc \\
4  &  -6.3998   & $T_{d}$       &     & 44 &  -122.5599 & $C_{2}$ & \\
5  &  -8.9663   & $D_{3h}$      &     & 45 &  -125.6996 & $C_{2v}$& fcc \\
6  &  -11.8950  & $O_{h}$       & fcc & 46 &  -128.5274 & $C_{2}$ & \\
7  &  -14.5508  & $D_{5h}$      &     & 47 &  -131.4723 & $C_{2v}$& \\
8  &  -17.2960  & $D_{2d}$      &     & 48 &  -134.5603 & $C_{2}$ & \\
9  &  -20.0965  & $D_{3h}$      &     & 49 &  -137.4842 & $C_{s}$ & \\
10 &  -22.8679  & $D_{4d}$      &     & 50 &  -140.8376 & $D_{3h}$& fcc \\
11 &  -25.5008  & $C_{2v}$      & cI  & 51 &  -143.7037 & $C_{3v}$& ucI \\
12 &  -28.5274  & $C_{5v}$      & cI  & 52 &  -146.9402 & $D_{2h}$& fcc \\
13 &  -32.0729  & $I_{h}$       & cI  & 53 &  -149.9979 & $C_{5v}$& ucI \\
14 &  -34.4434  & $C_{3v}$      & cI  & 54 &  -153.1459 & $I_{h}$ & ucI \\
15 &  -37.4486  & $D_{6d}$      &     & 55 &  -155.9151 & $I_{h}$ & cI \\
16 &  -40.2857  & $C_{2v}$      &     & 56 &  -158.6939 & $C_{1}$ & \\
17 &  -43.1633  & $D_{4h}$      & mix & 57 &  -161.8106 & $C_{s}$ & fcc \\
18 &  -45.8783  & $C_{4v}$      & mix & 58 &  -164.8037 & $C_{3v}$& \\
19 &  -48.8299  & $D_{5h}$      & cI  & 59 &  -167.8936 & $C_{1}$ & fcc \\
20 &  -51.7096  & $D_{2h}$      & fsI & 60 &  -170.8159 & $C_{2v}$& dec(5)\\
21 &  -54.5367  & $C_{s}$       & fsI & 61 &  -174.1955 & $C_{3v}$& fcc  \\
22 &  -57.5353  & $C_{s}$       & fsI & 62 &  -176.8996 & $C_{s}$ & fcc \\
23 &  -60.4193  & $C_{1}$       & fsI & 63 &  -179.9652 & $C_{s}$ & fcc \\
24 &  -63.2273  & $C_{s}$       & fsI & 64 &  -183.1181 & $C_{2v}$& dec(5)\\
25 &  -66.1897  & $C_{3}$       & fsI & 65 &  -186.0925 & $C_{2v}$& fcc \\
26 &  -69.0988  & $C_{1}$       & fsI & 66 &  -189.1802 & $C_{s}$ & fcc \\
27 &  -72.0921  & $C_{2v}$      & fsI & 67 &  -192.2851 & $C_{2v}$& dec(5)\\
28 &  -74.9678  & $C_{s}$       & fsI & 68 &  -195.5431 & $T_{d}$ & fcc  \\
29 &  -77.8530  & $C_{1}$       & fsI & 69 &  -198.2053 & $C_{1}$ & fcc \\
30 &  -80.8463  & $C_{s}$       & fsI & 70 &  -201.5432 & $C_{2v}$& fcc \\
31 &  -83.9112  & $C_{s}$       & fsI & 71 &  -204.6298 & $C_{s}$ & fcc \\
32 &  -86.8113  & $C_{2}$       & fsI & 72 &  -207.4224 & $C_{2v}$& dec(5) \\
33 &  -89.6630  & $C_{1}$       & fsI & 73 &  -210.6064 & $D_{5h}$& dec(5) \\
34 &  -92.7060  & $C_{1}$       & fsI & 74 &  -213.7521 & $C_{5v}$& dec(5) \\
35 &  -95.7977  & $D_{3}$       & fsI & 75 &  -216.9853 & $C_{s}$ & fcc \\
36 &  -98.6907  & $C_{2v}$      & fsI & 76 &  -219.6910 & $C_{4}$ & fcc \\
37 &  -101.6952 & $C_{3v}$      & fcc & 77 &  -222.8998 & $C_{s}$ & fcc \\
38 &  -105.1156 & $O_{h}$       & fcc & 78 &  -225.9885 & $C_{s}$ & fcc \\
39 &  -107.8211 & $C_{4v}$      & fcc & 79 &  -229.1335 & $D_{3h}$& fcc \\
40 &  -110.5958 & $C_{1}$       & ucI & 80 &  -231.9938 & $C_{4v}$& fcc \\
\end{tabular}
\end{ruledtabular}
\end{table}
\endgroup

\newpage

\begingroup
\squeezetable
\begin{table}
\caption{\label{table-Au}Global minima for Au clusters. For each
minimum energy ($E$), point group ($PG$) and structural assignment
($SA$) are given if possible. The structural categories are:
centred (cI), uncentred (ucI) and face-sharing icosahedral (fsI);
face centred cubic packed (fcc); decahedral with \textit{n} atoms
along the decahedral axis (dec(\textit{n})); involving a mixture
of staking sequences (mix).}
\begin{ruledtabular}
\begin{tabular}{l l l l l l l l}
$N$& $E$ (eV) & $PG$ & $SA$ & $N$ & $E$ (eV) & $PG$ & $SA$ \\
\hline
   &           &               & & 41 & -138.2008  & $C_{s}$   & fsI \\
2  &  -2.2886  & $D_{\infty h}$& & 42 & -141.9077  & $C_{4}$   & fsI \\
3  &  -5.2797  & $D_{3h}$&       & 43 & -145.4197  & $C_{s}$   &     \\
4  &  -8.8497  & $T_{d}$ &       & 44 & -149.0400  & $C_{s}$   &     \\
5  &  -12.1736 & $D_{3h}$&       & 45 & -152.6610  & $C_{s}$   &     \\
6  &  -15.8281 & $O_{h}$ & fcc   & 46 & -156.2059  & $C_{s}$   &     \\
7  &  -19.1505 & $D_{5h}$&       & 47 & -159.8067  & $C_{s}$   &     \\
8  &  -22.4326 & $D_{2d}$&       & 48 & -163.4242  & $C_{s}$   &     \\
9  &  -25.7507 & $C_{2v}$& cI    & 49 & -167.0450  & $C_{s}$   &     \\
10 &  -29.1712 & $C_{3v}$& cI    & 50 & -170.5777  & $C_{s}$   &     \\
11 &  -32.4968 & $C_{2v}$& cI    & 51 & -174.1304  & $C_{s}$   &     \\
12 &  -36.0088 & $C_{5v}$& cI    & 52 & -177.9191  & $C_{2h}$  & ucI \\
13 &  -40.1043 & $I_{h}$ & cI    & 53 & -181.7385  & $C_{5v}$  & ucI \\
14 &  -42.9943 & $C_{3v}$& cI    & 54 & -185.5635  & $I_{h}$   & ucI \\
15 &  -46.6960 & $D_{6d}$&       & 55 & -188.6971  & $C_{s}$   &     \\
16 &  -50.1275 & $C_{s}$ &       & 56 & -192.2661  & $C_{s}$   &     \\
17 &  -53.5914 & $C_{s}$ &       & 57 & -195.8573  & $C_{s}$   &     \\
18 &  -56.9242 & $C_{4v}$& mix   & 58 & -199.4305  & $C_{s}$   &     \\
19 &  -60.3352 & $D_{5h}$& cI    & 59 & -202.9304  & $C_{s}$   &     \\
20 &  -63.7463 & $C_{2h}$& fsI   & 60 & -206.6851  & $C_{s}$   &     \\
21 &  -67.2933 & $C_{s}$ & fsI   & 61 & -210.3464  & $C_{3v}$  & fcc \\
22 &  -70.9625 & $C_{s}$ & fsI   & 62 & -213.9025  & $C_{s}$   &     \\
23 &  -74.5236 & $C_{s}$ & fsI   & 63 & -217.5417  & $C_{2v}$  &     \\
24 &  -77.9539 & $C_{s}$ & fsI   & 64 & -221.2716  & $C_{2v}$  & dec(5) \\
25 &  -81.3036 & $C_{s}$ & fsI   & 65 & -224.9052  & $C_{s}$   &     \\
26 &  -84.8046 & $C_{s}$ & fsI   & 66 & -228.6560  & $C_{s}$   & mix \\
27 &  -88.4414 & $C_{s}$ & fsI   & 67 & -232.1324  & $C_{s}$   &     \\
28 &  -92.0749 & $C_{s}$ & fsI   & 68 & -235.8811  & $C_{s}$   &     \\
29 &  -95.5729 & $C_{s}$ & fsI   & 69 & -239.5284  & $C_{s}$   &     \\
30 &  -99.2318 & $C_{3v}$& fsI   & 70 & -243.1537  & $C_{s}$   &     \\
31 &  -102.5796& $C_{s}$ & fsI   & 71 & -246.8875  & $C_{2v}$  & dec(5) \\
32 &  -106.1560& $D_{2d}$& fsI   & 72 & -250.5921  & $C_{s}$   & mix \\
33 &  -109.6664& $C_{s}$ & fsI   & 73 & -254.3504  & $C_{s}$   & mix \\
34 &  -113.2711& $C_{s}$ & fsI   & 74 & -257.8233  & $C_{s}$   & mix \\
35 &  -116.8575& $C_{s}$ & fsI   & 75 & -261.6719  & $D_{5h}$  & dec(5)\\
36 &  -120.4893& $C_{2v}$& fsI   & 76 & -265.2637  & $C_{s}$   & mix \\
37 &  -124.0150& $C_{2v}$& mix   & 77 & -269.0221  & $C_{s}$   & mix \\
38 &  -127.6334& $O_{h}$ & fcc   & 78 & -272.4326  & $C_{s}$   &     \\
39 &  -131.1339& $C_{s}$ & fsI   & 79 & -276.1669  & $D_{3h}$  & fcc \\
40 &  -134.8451& $C_{s}$ & fsI   & 80 & -279.8123  & $C_{s}$   &     \\
\end{tabular}
\end{ruledtabular}
\end{table}
\endgroup

\newpage

\begingroup
\squeezetable
\begin{table}
\caption{\label{table-Pt}Global minima for Pt clusters. For each
minimum energy ($E$), point group ($PG$) and structural assignment
($SA$) are given if possible. The structural categories are:
centred (cI), uncentred (ucI) and face-sharing icosahedral (fsI);
face centred cubic packed (fcc); decahedral with \textit{n} atoms
along the decahedral axis (dec(\textit{n})); involving a mixture
of staking sequences (mix).}
\begin{ruledtabular}
\begin{tabular}{l l l l l l l l}
$N$& $E$ (eV) & $PG$ & $SA$ & $N$ & $E$ (eV) & $PG$ & $SA$ \\
\hline
   &            &               & & 41 & -199.6745& $C_{s}$& mix    \\
2  &  -3.1515   & $D_{\infty h}$& & 42 & -204.9297& $C_{4}$& fsI    \\
3  &  -7.3640   & $D_{3h}$&       & 43 & -210.1853& $C_{2}$& cI     \\
4  &  -12.4627  & $T_{d}$ &       & 44 & -215.5259& $C_{s}$& cI     \\
5  &  -17.2131  & $D_{3h}$&       & 45 & -220.6938& $C_{s}$&        \\
6  &  -22.4353  & $O_{h}$ & fcc   & 46 & -225.9648& $C_{s}$& cI     \\
7  &  -27.2189  & $D_{5h}$&       & 47 & -231.1459& $C_{s}$& cI     \\
8  &  -31.8884  & $D_{2d}$&       & 48 & -236.3969& $C_{s}$& cI     \\
9  &  -36.7091  & $C_{2v}$& cI    & 49 & -241.7241& $C_{s}$& cI     \\
10 &  -41.6455  & $C_{3v}$& cI    & 50 & -246.8295& $D_{3h}$& fcc   \\
11 &  -46.4621  & $C_{2v}$& cI    & 51 & -252.1407& $C_{3v}$& ucI   \\
12 &  -51.6089  & $C_{5v}$& cI    & 52 & -257.7687& $C_{2h}$& ucI   \\
13 &  -57.5826  & $I_{h}$ & cI    & 53 & -263.3864& $C_{5v}$& ucI   \\
14 &  -61.7317  & $C_{3v}$& cI    & 54 & -269.0105& $I_{h}$ & ucI   \\
15 &  -66.9514  & $D_{6d}$&       & 55 & -273.4541& $I_{h}$ & cI    \\
16 &  -71.8609  & $C_{2v}$&       & 56 & -278.3894& $C_{s}$ &       \\
17 &  -76.8300  & $C_{s}$ &       & 57 & -283.6149& $C_{s}$ &       \\
18 &  -81.6960  & $C_{2v}$&       & 58 & -288.8067& $C_{s}$ &       \\
19 &  -86.9222  & $D_{5h}$& cI    & 59 & -293.9930& $C_{s}$ &       \\
20 &  -91.7288  & $C_{2h}$& fsI   & 60 & -299.3716& $C_{s}$ &       \\
21 &  -96.8290  & $C_{s}$ & fsI   & 61 & -304.8093& $C_{3v}$& fcc  \\
22 &  -102.0877 & $C_{s}$ & fsI   & 62 & -310.0821& $C_{s}$ &       \\
23 &  -107.2310 & $C_{s}$ & fsI   & 63 & -315.4525& $C_{2v}$&       \\
24 &  -112.1612 & $C_{s}$ & fsI   & 64 & -320.8369& $C_{2v}$& dec(5)\\
25 &  -117.0114 & $C_{3}$ & fsI   & 65 & -325.9583& $C_{s}$ &       \\
26 &  -122.2412 & $C_{s}$ & fsI   & 66 & -331.2959& $C_{s}$ & mix \\
27 &  -127.4586 & $C_{s}$ & fsI   & 67 & -336.4653& $C_{s}$ &     \\
28 &  -132.7066 & $C_{s}$ & fsI   & 68 & -341.9533& $C_{s}$ &     \\
29 &  -137.7405 & $C_{2}$ & fsI   & 69 & -347.1607& $C_{s}$&     \\
30 &  -143.0386 & $C_{3v}$& fsI   & 70 & -352.4811& $C_{s}$ & dec(5)\\
31 &  -147.9993 & $C_{3}$ & fsI   & 71 & -358.1813& $C_{2v}$& dec(5) \\
32 &  -153.1794 & $D_{2d}$& fsI   & 72 & -363.2608& $C_{s}$ & mix \\
33 &  -158.2298 & $C_{2}$ & fsI   & 73 & -368.7448& $C_{s}$ & mix \\
34 &  -163.3569 & $C_{s}$ & fsI   & 74 & -374.0287& $C_{5v}$& dec(5)    \\
35 &  -168.7294 & $D_{3}$ & fsI   & 75 & -379.7413& $D_{5h}$& dec(5)    \\
36 &  -173.9244 & $C_{2v}$& fsI   & 76 & -384.6942& $C_{2v}$& dec(5)    \\
37 &  -179.0675 & $C_{2v}$& mix   & 77 & -390.1332& $C_{2v}$& mix \\
38 &  -184.4825 & $O_{h}$ & fcc   & 78 & -395.0530& $C_{s} $& mix \\
39 &  -189.4859 & $C_{s}$ & cI    & 79 & -400.7173& $D_{3h}$& fcc  \\
40 &  -194.8158 & $D_{2}$ & fsI   & 80 & -405.7957& $C_{s}$ & mix  \\
\end{tabular}
\end{ruledtabular}
\end{table}
\endgroup

\newpage


\newpage

\begin{thebibliography}{abdx}

\bibitem{Feynman}R.P. Feynman, Talk at the annual meeting of the American
Physical Society at the California Institute of Technology
(Caltech), December $29^{th}$, (1959).

\bibitem{Haberland} H. Haberland (Ed.), {\it Clusters of Atoms
and Molecules} (Springer, Berlin, 1994); and references therein.

\bibitem{Schmid} G. Schmid (Ed.), {\it Clusters and Colloids} (VCH,
Weinheim, 1994); and references therein.

\bibitem{Martin} T.P. Martin (Ed.), {\it Large Clusters of Atoms
and Molecules} (Kluwer, Dordrecht, 1996); and references therein.

\bibitem{Jellinek} J. Jellinek (Ed.), {\it Theory of Atomic and
Molecular Clusters} (Springer, Berlin, 1999); and references
therein.

\bibitem{Johnston} Roy L. Johnston, {\it Atomic and Molecular Clusters}
(Taylor and Francis, London, 2002); and references therein.

\bibitem{Eberhardt}W. Eberhardt, Surf. Sci. 500, 242 (2002).

\bibitem{Jel-Guv} J. Jellinek and Z.B. G\"{u}ven\c{c},
Z. Phys. D 26, 110 (1993); J. Jellinek and Z.B. G\"{u}ven\c{c}, in
{\it The Synergy Between Dynamics and Reactivity at Clusters and
Surfaces} (L.J. Farrugia, Ed. Kluwer, Dordrecht, 1995, p.217).

\bibitem{Wales} D.J. Wales and J.P.K. Doye, J. Phys. Chem. A 101, 5111
(1997).

\bibitem{Voter}A.F. Voter, Los Alamos Unclassified Technical Report \#LA-UR
93-3901 (1993).

\bibitem{Baskes}M.S. Daw and M.I. Baskes, Phys. Rev. B 29, 6443 (1984).

\bibitem{Goldberg}D.E. Goldberg, {\it Genetic Algorithms in Search,
Optimisation and Machine Learning} (Addison-Wesley, Reading, MA,
1989).

\bibitem{Stillinger}F.H. Stillinger and T.A. Weber, J. Stat. Phys.
52, 1429 (1988).

\bibitem{Doye1}J.P.K. Doye, and D.J. Wales, New J. Chem. 733 (1998).

\bibitem{Doye2}J.P.K. Doye, Phys. Rev. B 68(19), 195418 (2003); and references therein.

\bibitem{Li}Z. Li and H.A. Scheraga, Proc. Natl. Acad. Sci. USA,
84, 6611 (1987).

\bibitem{opt}http://www-wales.ch.cam.ac.uk/software.html.

\bibitem{Doye3}J.P.K. Doye, D.J. Wales, and R.S. Berry, J. Chem. Phys. 103, 4234 (1995).

\bibitem{Doye4}J.P.K. Doye and D.J. Wales, J. Chem. Soc. Faraday Trans. 93, 4233 (1997).

\bibitem{Chou}M.Y. Chou and M.L. Cohen, Phys. Lett. A 113, 420 (1986).

\bibitem{Jug}K. Jug, H.P. Schluff, H. Kupka, and R. Iffert, J.
Comput. Chem. 9, 803 (1988).

\bibitem{Pacchioni}G. Pacchioni and J. Koutecky, Ber. Bunsenges.
Phys. Chem. 88, 242 (1984).

\bibitem{Upton}T.H. Upton, J. Phys. Chem. 90, 754, (1986); Phys.
Rev. Lett. 56, 2168 (1986).

\bibitem{Pettersson}L.G.M. Petersson, C.W. Bauschlicher, Jr., and
T. Halicioglu, J. Chem. Phys. 87, 2205 (1987).

\bibitem{Cheng}H.P. Cheng, R.S. Berry, and R.L. Whetten, Phys.
Rev. B 43, 10647 (1991).

\bibitem{Yi}J.Y. Yi, D.J. Oh, and J. Bernhole, Phys. Rev. Lett.
67, 1594 (1991).

\bibitem{Jones}R.O. Jones, Phys. Rev. Lett. 67, 224, (1991); J.
Chem. Phys. 99, 1194 (1993).

\bibitem{Akola}J. Akola, H. Hakkinen, and M. Manninen, Phys. Rev.
B 58, 3601 (1998).

\bibitem{Ahlriches}R. Ahlrichs and S.D. Elliott, Phys. Chem. Chem.
Phys. 1, 13 (1999).

\bibitem{Khanna}S.N. Khanna and P. Jena, Phys. Rev. Lett. 69, 1664 (1992).

\bibitem{Gong}X.G. Gong and V. Kumar, Phys. Rev. Lett. 70, 2078 (1993).

\bibitem{Jellinek2}E.B. Krissinel and J. Jellinek, Int. J. Quantum
Chem. 62, 185 (1997).

\bibitem{Rao}B.K. Rao and P. Jena, J. Chem. Phys. 111, 1890 (1999).

\bibitem{Erkoc}Z. El-Bayyari and \c{S}. Erko\c{c}, Phys. Status
Solidi B 170, 103 (1992).

\bibitem{Johnston2}R.L. Johnston and J.-Y. Fang, J. Chem. Phys. 97,
7809 (1992)

\bibitem{Lloyd1}L.D. Lloyd and R.L. Johnston, Chem. Phys. 236, 107
(1998).

\bibitem{Lloyd2}L.D. Lloyd, R.L. Johnston, C. Roberts, and T.V. Mortimer-Jones,
Chem. Phys. Chem. 3, 408 (2002).

\bibitem{Turner}G.W. Turner, R.L. Johnston, and N.T. Wilson, J.
Chem. Phys. 112, 4773 (1999).

\bibitem{Joswig}J.-O. Joswig and M. Springborg, Phys. Rev. B 68,
085408 (2003).

\bibitem{Cox}D.M. Cox, D.J. Trevor, R.L. Whetten, E.A. Rohlfing,
and A. Kaldor, J. Chem. Phys. 84, 4651 (1986).

\bibitem{Jarrold}M.F. Jarrold, J.E. Bower, and J.S. Kraus, J.
Chem. Phys. 86, 3876 (1987).

\bibitem{Hanley}L. Hanley, S. Ruatta, and S. Anderson, J. Chem.
Phys. 87, 260 (1987).

\bibitem{Saunders}W.A. Saunders, P. Fayet, and L. W\"{o}te, Phys.
Rev. A 39, 4400 (1989).

\bibitem{Leuchtner}R.E. Leuchtner, A.C. Harms, and A.W. Castleman,
Jr., J. Chem. Phys. 91, 2753 (1989); 94, 1093 (1991).

\bibitem{Schriver}K.E. Schriver, J.L. Persson, E.C. Honea, and
R.L. Whetten, Phys. Rev. Lett. 64, 2539 (1990).

\bibitem{Heer}W.A. de Heer, P. Milani, and A. Chatelain, Phys. Rev
Lett. 63, 2834 (1989).

\bibitem{Gantefor}G. Gantef\"{o}r, M. Gausa, K.H. Meiwes-Broer,
and H.O. Lutz, Z. Phys. D 9, 253 (1988).

\bibitem{Taylor}K.J. Taylor, C.L. Pettiette, M.J. Graycraft, O.
Chesnovsky, and R.E. Smalley, Chem. Phys. Lett. 152, 347 (1988).

\bibitem{Nakajima}A. Nakajima, K. Hoshino, T. Naganuma, Y. Sone,
and K. Kaya, J. Chem. Phys. 95, 7061 (1991).

\bibitem{Wu}X. Li, H. Wu, X.B. Wang, and L.S. Wang, Phys. Rev.
Lett. 81, 1090 (1998).

\bibitem{Martin2} T.P. Martin, Phys. Rep. 199, 273 (1996).

\bibitem{Knight}W.D. Knight, K. Clemenger, W.A. de Heer, W.A.
Saunders, M.Y. Chou, M.L. Cohen, Phys. Rev. Lett. 52, 2141 (1984).

\bibitem{Northby} J.A. Northby, J. Xie, D.L. Freeman,J.D. Doll, Z. Phys. D 12, 69 (1989).

\bibitem{Stein} J.W. Lee, G.D. Stein, J. Phys. Chem. 91, 2450 (1987).

\bibitem{Clemenger} K. Clemenger, Phys. Rev. B 32, 1359 (1985).

\bibitem{Ali1}A. Sebetci and Z.B. G\"{u}ven\c{c}, Surf. Sci.
525, 66 (2003).

\bibitem{Mackay}A.L.Mackay, Acta Crystallogr. 15, 916 (1962).

\bibitem{Ali2}A. Sebetci and Z.B. G\"{u}ven\c{c}, Eur. Phys. J. D,
30(1), 71 (2004).

\bibitem{Whetten}R.L. Whetten, M.N. Shafigullin, J.T. Khoury, T.G. Schaaff,
I. Vezmar, M.M. Alvarez, A. Wilkinson, Acc. Chem. Res. 32(5), 397
(1999).

\bibitem{Mirkin}J.-M. Nam, C.S. Thaxton, C.A. Mirkin,
Science 301, 1884 (2003).

\bibitem{Andres}J Liu, T Lee, D.B. Janes, B.L. Walsh, M.R. Melloch,
J.M. Woodall, R. Reifenberger, R.P. Andres, Appl. Phys. Lett.
77(3), 373 (2000).

\bibitem{Cleveland}R.N. Barnett, C.L. Cleveland, H. Hakkinen, W.D. Luedtke,
C. Yannouleas C, U Landman, Eur. Phys. J. D 9(1-4), 95 (1999).

\bibitem{Schaaff}T.G. Schaaff, M.N. Shafigullin, J.T. Khoury,
I. Vezmar, R.L. Whetten, W.G. Cullen, P.N. First, C.
GutierrezWing, J. Ascensio, M.J. JoseYacaman, J. Phys. Chem. B
101, 7885 (1997).

\bibitem{Koga}K. Koga, H. Takeo, T. Ikeda, K.I. Ohshima, Phys.
Rev. B 57, 4053 (1998).

\bibitem{Palpant}B. Palpant, B. Prevel, J. Lerme, E. Cottancin, M. Pellarin,
M. Treilleux, A. Perez, J.L. Vialle, M. Broyer, Phys. Rev. B 57,
1963 (1998).

\bibitem{Spasov}V.A. Spasov, Y. Shi, K.M. Ervin, Chem. Phys.
262, 75 (2000).

\bibitem{Luedtke}C.L. Cleveland, W.D. Luedtke, U. Landman,
Phys. Rev. B 60(7), 5065 (1999).

\bibitem{Garzon}I.L. Garzon, K. Michealian, M.R. Beltran, A. Posada-Amarillas,
P. Ordejon, E. Artacho, D. Sanchez-Portal, J.M. Soler,  Phys. Rev.
Lett. 81, 1600 (1998).

\bibitem{Wilson}N.T. Wilson and R.L. Johnston, Eur. Phys. J. D 12,
161 (2000).

\bibitem{Haberlen}O. D. H\"{a}berlen, S.-C. Chung, M. Stener, N.
Rösch, J. Chem. Phys. 106, 5189 (1997).

\bibitem{Wang}J.L. Wang, G.H. Wang, J.J. Zhao, Phys.
Rev. B 66(3), Art. No. 035418 (2002).

\bibitem{Hakkinen}H. H\"{a}kkinen and U. Landman, Phys.
Rev. B 62, 2287 (2000).

\bibitem{Gronbech}H. Gr\"{o}nbech and W. Andreoni, Chem, Phys.
262, 1 (2000).

\bibitem{Bravo}G. Bravo-Perez, I.L. Garzon, O. Novaro, J. Mol. Struct.:
THEOCHEM 493, 225 (1999).



\end{thebibliography}
\end{document}